\documentclass[12pt,a4paper,twoside]{article}
\usepackage{graphics}
\usepackage{delarray}
\newlength{\figwidth}
\newlength{\figheight}
\newcommand{\bigint}[2]{\displaystyle \int\limits_{#1}^{#2}} 
\newcommand{\bigmod}[2]{\left.{#1}\strut\right|_{#2}}
\newcommand{\eqref}[1]{(\ref{#1})}
\title{ 
\vspace{-2cm}
  {\large\sc   Institute of Experimental Physics} \\[-6mm]
  {\large \sc  Warsaw University} 
\vspace{-0.5cm}
\begin{flushright}
\begin{tabular}{l}
 {\large IFD - 01/2000} \\[-3mm]
 {\large hep-ph/0003271 } 
\end{tabular}
\end{flushright}
\vspace{0.5cm}
Leptoquark signal from global analysis } 
\author{Aleksander Filip \.Zarnecki \\
{\small\it 
Institute of Experimental Physics, 
Warsaw University,
Ho\.za 69, 00-681 Warszawa, Poland} \\
{\small\it E-mail: zarnecki@fuw.edu.pl}
} 
\date{April 30, 2000} 
\begin{document} 

\maketitle 

\begin{abstract}
Data from HERA, LEP and the Tevatron, as well as from low energy  
experiments are used to constrain the Yukawa couplings
for scalar and vector leptoquarks
in the Buchm\"uller-R\"uckl-Wyler effective model.
In the limit of very high leptoquark masses constraints on the coupling
to the mass ratio $\lambda / M$ are derived using 
the contact-interaction approximation.
For finite masses the coupling limits are studied 
as a function of the leptoquark mass.
Some leptoquark models are found to describe the existing experimental data 
much better than the Standard Model.
Increase in the global probability observed for models including 
$S_{1}$ or $\tilde{V}_{\circ}$
leptoquark production/exchange  
corresponds to more than a 3$\sigma$ effect.
Assuming that a real leptoquark signal is observed,
calculated is an allowed region in the $\lambda - M$ plane.
The leptoquark signal is mostly resulting from the new data 
on the atomic parity violation in cesium,
but is also supported by recent LEP2 measurements,
unitarity violation in the CKM matrix
and  HERA high-$Q^{2}$ results.
\end{abstract}

\thispagestyle{empty}

%
%

\section{Introduction}
\label{sec-intro}

In 1997 the H1 \cite{h1_97} and ZEUS \cite{zeus97} experiments at HERA 
reported  an excess of events in positron-proton Neutral Current Deep 
Inelastic Scattering (NC DIS) at very high momentum transfer 
scales $Q^{2}$, as compared with the predictions of the Standard Model. 
As a possible sign of some "new physics" this results provoked many
theoretical speculations.
Clustering of H1 events at a positron-jet invariant masses of about 200 GeV
was considered to indicate a possible resonant production of leptoquark states.
The agreement with the Standard Model prediction improved after both 
experiments doubled their positron-proton data samples, but
some discrepancy is still there  
and calls for a better understanding.

In a  paper presented last year \cite{mygcia},\footnote{For recent update of
presented results see \cite{dis2000}.} data from
HERA, LEP and the Tevatron, as well as from low energy  
experiments were used to constrain the mass scale of 
the possible new electron-quark contact interactions. 
A contact interaction model was used as the most general framework
which can describe possible low energy effects coming from "new physics" 
at much higher energy scales. 
This includes the possible existence of second-generation 
heavy weak bosons, leptoquarks, as well as electron and quark 
compositeness \cite{cidef,cihera}.
In addition to the general models, in which all new contact interaction
couplings can vary independently, the global analysis considered
also a set of one-parameter models which assumed fixed relations between 
couplings.
However, only parity conserving models were selected, 
as suggested by ZEUS \cite{zeusci}, to avoid strong limits
coming from atomic parity violation (APV) measurements \cite{apvold}.
No significant improvement in the description of the data 
has been obtained for any of these models.

Theoretical uncertainties in the parity violation measurements
in cesium atoms have been recently significantly reduced.
As a result, measured value of the cesium weak charge is now more than
2$\sigma$ away from the Standard Model predictions \cite{apvnew}.
This discrepancy could be due to new parity-violating electron-quark
interactions.
Considered in this paper are effects induced by the possible existence 
of the first-generation leptoquarks.
Predictions based on the Buchm\"uller-R\"uckl-Wyler (BRW) effective 
model \cite{brw} are compared with the existing experimental data. 
In the limit of very high masses, the exchange of leptoquarks 
can be described using the contact interaction approach \cite{lqci}.
Limits on the ratio of the coupling and the mass are derived.
For finite leptoquark masses 
limits on leptoquark Yukawa coupling $\lambda$ are
studied as a function of the leptoquark mass.

The aim of the present analysis is to combine the APV measurements with
other data to constrain leptoquark coupling and mass, 
and to look for a possible leptoquark signal in the combined data.
The BRW model used in this
analysis is described in  section \ref{sec-model}. 
In section \ref{sec-data} the relevant data 
from HERA, LEP, the Tevatron and other experiments are briefly 
described. 
Methods used to compare data with leptoquark
model predictions and to derive coupling limits
are summarised in section \ref{sec-method}.
The analysis results for different leptoquark types,
including extracted coupling-mass limits
and discussion of the possible leptoquark signal 
are presented in section \ref{sec-results}.

The analysis presented here is based on the approach used in
the global analysis of $eeqq$ contact interactions \cite{mygcia,dis2000},
which in turn followed \cite{ciglob,cile}.
When finalising this analysis another work discussing leptoquark
exchange as a possible explanation for the APV result 
was released \cite{apvlq}.
However, the analysis presented there is limited to 
the contact interaction approximation.


\section{Leptoquark models}
\label{sec-model}

Striking symmetry between quarks and leptons in the Standard Model
strongly suggests that, if there exist a more fundamental theory
it should also introduce a more fundamental relation between them.
Such lepton-quark "unification" is achieved for example 
in different theories of grand unification \cite{gut} 
and in compositeness models.
Whenever quarks and leptons are allowed to couple directly 
to each other, a quark-lepton bound state can also exist.
Such particles, called leptoquarks, carry both colour and 
fractional electric charge and a lepton number.
Also supersymmetric theories with broken R-parity predict
squarks (leptoquark type objects) coupling to quark-lepton pairs.

In this paper a general classification of leptoquark states 
proposed by Buchm\"uller, R\"uckl and Wyler \cite{brw} will be used.
The Buchm\"uller-R\"uckl-Wyler (BRW) model is based on 
the assumption that new interactions should respect the 
$SU(3)_{C} \times SU(2)_{L} \times U(1)_{Y}$ symmetry of
the Standard Model.
In addition leptoquark couplings are assumed to be family diagonal
(to avoid FCNC processes) and to conserve lepton and baryon numbers
(to avoid rapid proton decay).
Taking into account very strong bounds from rare decays
it is also assumed that leptoquarks couple either to left-
or to right-handed leptons.
With all these assumptions there are 14 possible states 
(isospin singlets or multiplets) of scalar and vector leptoquarks.
Table \ref{tab-aachen} lists these states according to 
the so-called Aachen notation \cite{aachen}.
An S(V) denotes a scalar(vector) leptoquark and the subscript
denotes the weak isospin.
When the leptoquark can couple to both right- and left-handed
leptons, an additional superscript indicates  the lepton chirality.
A tilde is introduced to differentiate between leptoquarks
with different hypercharge.
\begin{table}[tbp]
  \begin{center}
   \begin{tabular}{lcccccc}
      \hline\hline\hline\noalign{\smallskip}
Model & Fermion & Charge & $BR(LQ \rightarrow e^{\pm}q)$ & 
        \multicolumn{2}{c}{Coupling} & Squark \\
      & number F &   Q   & $\beta$  &   &  & type \\
\hline\hline\hline\noalign{\smallskip}
$S_{\circ}^L$ &  2  &  $-1/3$  &  1/2  &   $e_{L}u$ & $\nu d$  & $\tilde{d_R}$ \\
\hline\noalign{\smallskip}
$S_{\circ}^R$ &  2  &  $-1/3$  &  1  &  $e_{R}u$ &  &   \\
\hline\noalign{\smallskip}
$\tilde{S}_{\circ}$  &  2  &  $-4/3$   &  1  &  $e_{R}d$ & &    \\
\hline\noalign{\smallskip}
$S_{1/2}^L$   &  0  &  $-5/3$   &  1  &  $e_{L} \bar{u}$ & &  \\
              &     &  $-2/3$   &  0  &  & $\nu \bar{u}$  &   \\
\hline\noalign{\smallskip}
$S_{1/2}^R$   &  0  &  $-5/3$  &  1 &  $e_{R} \bar{u}$  &  &  \\
              &     &  $-2/3$  &  1 &  $e_{R} \bar{d}$  &  &  \\
\hline\noalign{\smallskip}
$\tilde{S}_{1/2}$ &  0 &  $-2/3$  &  1  &  $e_{L} \bar{d}$  &  & $\overline{\tilde{u}_{L}}$  \\
                  &    &  $+1/3$  &  0  &     & $\nu \bar{d}$  & $\overline{\tilde{d}_{L}}$  \\
\hline\noalign{\smallskip}
$S_{1}$       &  2  & $-4/3$  &  1  & $e_{L}d$ & & \\
              &     & $-1/3$  &  1/2 & $e_{L}u$ &  $\nu d$  & \\
              &     & $+2/3$  &  0   &  &  $\nu d$  &  \\
\hline\hline\hline\noalign{\smallskip}
$V_{\circ}^L$ &  0  &  $-2/3$  &  1/2  &   $e_{L}\bar{d}$ & $\nu \bar{u}$ & \\
\hline\noalign{\smallskip}
$V_{\circ}^R$ &  0  &  $-2/3$  &  1  &  $e_{R}\bar{d}$ & & \\
\hline\noalign{\smallskip}
$\tilde{V}_{\circ}$  &  0  &  $-5/3$   &  1  &  $e_{R}\bar{u}$  & &  \\
\hline\noalign{\smallskip}
$V_{1/2}^L$   &  2  &  $-4/3$   &  1  &  $e_{L} d$  & & \\
              &     &  $-1/3$   &  0  &  & $\nu d$  &   \\
\hline\noalign{\smallskip}
$V_{1/2}^R$   &  2  &  $-4/3$  &  1 &  $e_{R} d$  &  &   \\
              &     &  $-1/3$  &  1 &  $e_{R} u$  &  &  \\
\hline\noalign{\smallskip}
$\tilde{V}_{1/2}$ &  2 &  $-1/3$  &  1  &  $e_{L} u$ &  &  \\
                 &     &  $+2/3$   &  0  &  & $\nu u$  &   \\
\hline\noalign{\smallskip}
$V_{1}$       &  0  & $-5/3$  &  1  & $e_{L} \bar{u}$ & \\
              &     & $-2/3$  &  1/2 & $e_{L}\bar{d}$ & $\nu \bar{u}$ & \\
              &     & $+1/3$  &  0   &   &  $\nu \bar{d}$  &  \\
\hline\hline\hline\noalign{\smallskip}
    \end{tabular}
  \end{center}
  \caption{A general classification of leptoquark states 
 in the Buchm\"uller-R\"uckl-Wyler model. 
 Listed are the leptoquark fermion number, F, 
 electric charge, Q (in units of elementary charge), 
the branching ratio to electron-quark (or electron-antiquark), $\beta$
 and the flavours of the coupled lepton-quark pairs. 
Also shown are possible squark assignments to the leptoquark states
 in  the minimal supersymmetric theories with broken R-parity.}
  \label{tab-aachen}
\end{table}
Listed in Table \ref{tab-aachen} are the leptoquark fermion
number F, electric charge Q, and the branching ratio to an electron-quark
pair (or electron-antiquark pair), $\beta$.
The leptoquark branching fractions are predicted by the BRW model 
and are either 1, $\frac{1}{2}$ or 0.
For a given electron-quark branching ratio $\beta$, the branching ratio 
to the neutrino-quark is by definition $(1-\beta)$. 
Also included in Table \ref{tab-aachen} are the flavours and chiralities of 
the lepton-quark pairs coupling to a given leptoquark type.
In three cases the squark flavours (in supersymmetric theories with 
broken R-parity) with corresponding  couplings
are also indicated.
Present analysis takes into account only leptoquarks which couple
to the first-generation leptons ($e$, $\nu_{e}$) and first-generation 
quarks ($u$, $d$), as most of the existing experimental data  
constrain this type of couplings.
Second- and third-generation leptoquarks as well as generation-mixing
leptoquarks will not be considered in this paper.
It is also assumed that one of the leptoquark 
types gives the dominant contribution, as compared with other leptoquark
states 
and that the interference between different leptoquark states can be neglected.
Using this simplifying assumption, different leptoquark types 
can be considered separately.
Finally, it is assumed that different leptoquark states within 
isospin doublets and triplets have the same mass.

The $ep$ collider HERA is the unique place to search for the first-generation
leptoquarks, as single leptoquarks can directly be produced in
electron-quark interactions.
The influence of the leptoquark production or exchange on the $ep$ NC DIS 
cross-section can be described as an additional term in the tree level 
$eq \rightarrow eq$ scattering amplitude:\footnote{
Amplitude given for electron-quark scattering describes also 
scattering of positrons and anti-quarks taken with opposite chiralities.}
\begin{eqnarray}
M^{eq}_{ij}(s,t,u) & = & 
- \frac{4 \pi \alpha_{em} e_{q}}{t} \; + \;
  \frac{4 \pi \alpha_{em}}{\sin^{2}\theta_{W} \; \cos^{2}\theta_{W} } 
\cdot \frac{g^{e}_{i} g^{q}_{j}}{t - M_{Z}^{2}}
  \; + \; \eta^{eq}_{ij}(s,u) \;\; , \label{eq-mt}
\end{eqnarray}
where $s$, $t$ and $u$ are the Mandelstam variables describing the
electron-quark scattering subprocess, $e_{q}$ is 
the electric charge of the quark in units of the elementary charge,
the subscripts $i$ and $j$ label the chiralities of the initial 
lepton and quark, respectively  ($i,j=L,R$),
and $g^{e}_{i}$ and $g^{q}_{j}$ are electroweak 
couplings of the electron and the quark. 
In the limit $M_{LQ} \gg \sqrt{s}$  the leptoquark contribution
to the scattering amplitude given by $\eta^{eq}_{ij}(s,u)$ does not depend
on the process kinematics and can be written as
\begin{eqnarray}
\eta^{eq}_{ij} & = & 
a^{eq}_{ij} \cdot \left( \frac{\lambda_{LQ}}{M_{LQ}} \right)^{2} \;\; ,
\label{eq-cieta}
\end{eqnarray}
where $M_{LQ}$ is the leptoquark mass, $\lambda_{LQ}$ the leptoquark-electron-quark
Yukawa coupling and  the coefficients $a^{eq}_{ij}$ are 
given in Table \ref{tab-lqci} \cite{lqci}.
The effect of heavy leptoquark production or exchange
is equivalent to a vector type $eeqq$ contact interaction.
It is interesting to notice that 5 scalar leptoquark types
($S_{\circ}^R$, $\tilde{S}_{\circ}$, $S_{1/2}^L$, $S_{1/2}^R$ 
and $\tilde{S}_{1/2}$) correspond to the same contact interaction
coupling structures (but opposite coupling signs) as 5 vector models 
($\tilde{V}_{\circ}$, $V_{\circ}^R$, $\tilde{V}_{1/2}$, 
$V^{R}_{1/2}$ and $V^{L}_{1/2}$ respectively). 
\begin{table}[tbp]
  \begin{center}
   \begin{tabular}{lrrrrrrrr}
      \hline\hline\hline\noalign{\smallskip}
  Model & 
$a^{ed}_{LL}$ & $a^{ed}_{LR}$ & $a^{ed}_{RL}$ & $a^{ed}_{RR}$ & 
$a^{eu}_{LL}$ & $a^{eu}_{LR}$ & $a^{eu}_{RL}$ & $a^{eu}_{RR}$ \\ 
\noalign{\smallskip}
\hline\hline\hline\noalign{\smallskip}
$S_{\circ}^L$ &    &    &    &    & +$\frac{1}{2}$   &    &    &    \\
$S_{\circ}^R$ &    &    &    &    &    &    &    &  +$\frac{1}{2}$     \\
$\tilde{S}_{\circ}$ 
              &    &    &    &  +$\frac{1}{2}$     &    &    &    &    \\
$S_{1/2}^L$   &    &    &    &    &    &  $-\frac{1}{2}$     &    &    \\
$S_{1/2}^R$   &    &    &  $-\frac{1}{2}$ &   &   &   & $-\frac{1}{2}$ &    \\
$\tilde{S}_{1/2}$ 
              &    &  $-\frac{1}{2}$  &    &    &    &    &    &    \\
$S_{1}$       &  +1  &    &    &    &  +$\frac{1}{2}$   &    &    &    \\
\noalign{\smallskip}
\hline\noalign{\smallskip}
$V_{\circ}^L$ &  $-1$  &    &    &    &    &    &    &    \\
$V_{\circ}^R$ &    &    &    &  $-1$  &    &    &    &    \\
$\tilde{V}_{\circ}$ 
              &    &    &    &    &    &    &    &  $-1$   \\
$V_{1/2}^L$   &    &  +1  &    &    &    &    &    &    \\
$V_{1/2}^R$   &    &    &  +1  &    &    &    &  +1   &    \\
$\tilde{V}_{1/2}$ 
              &    &    &    &    &    &  +1  &    &    \\
$V_{1}$       & $-1$   &    &    &    &  $-2$  &    &    &    \\
\noalign{\smallskip}
\hline\hline\hline\noalign{\smallskip}
    \end{tabular}
  \end{center}
  \caption{Coefficients $a^{eq}_{ij}$ defining the effective 
           contact interaction couplings 
 $\eta^{eq}_{ij}=a^{eq}_{ij}\cdot \frac{\lambda_{LQ}^2}{M_{LQ}^2}$ 
 for different models of scalar (upper part of the table)
 and vector (lower part) leptoquarks.           
 Empty places in the table correspond to $a^{eq}_{ij}=0$.}
  \label{tab-lqci}
\end{table}

\begin{figure}[tbp]
\centerline{\resizebox{\figwidth}{!}{%
  \includegraphics{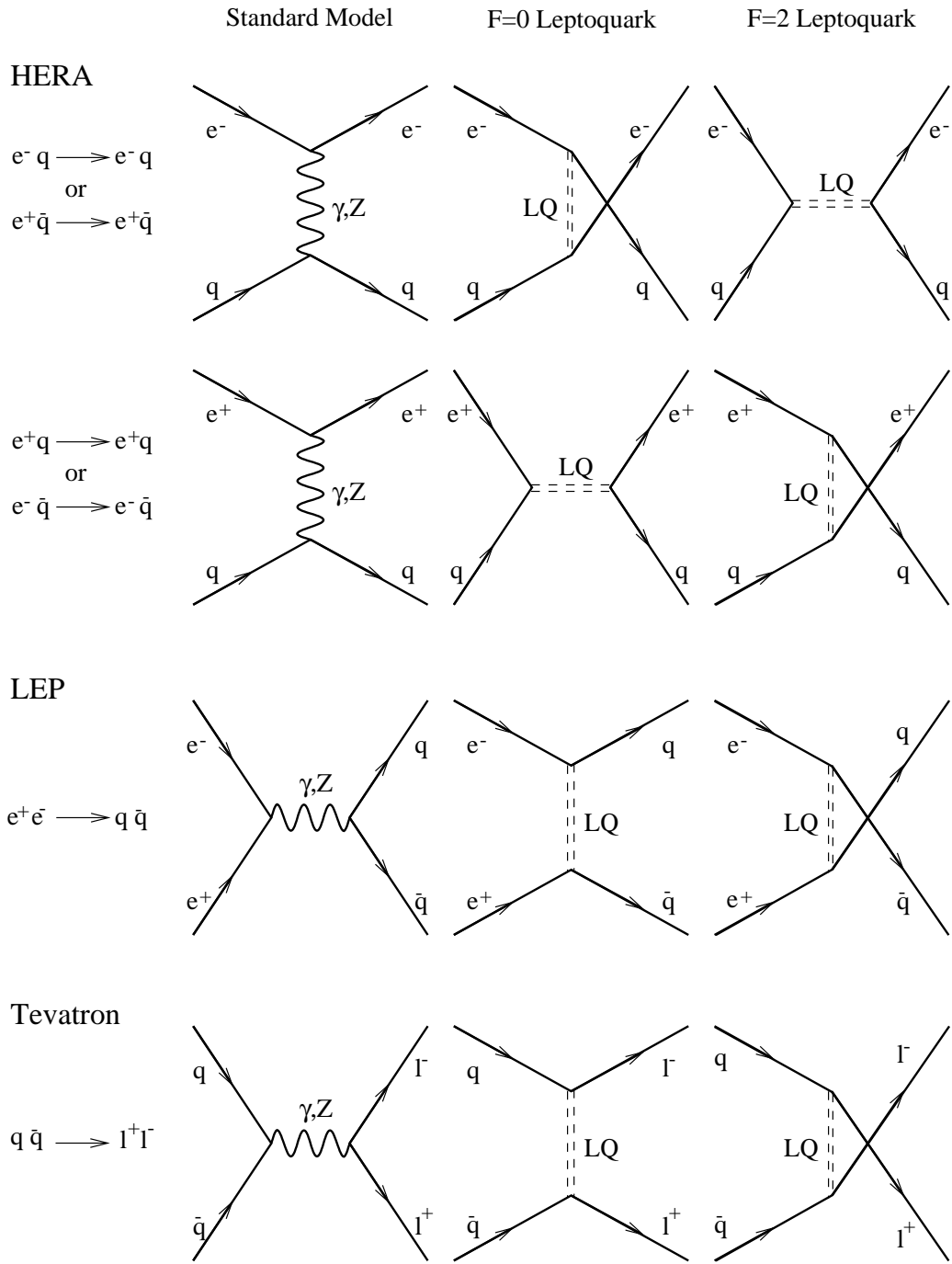}
}}
  \caption{Diagrams describing leading order Standard Model processes
           and leptoquark contributions coming from F=0 and F=2 leptoquarks,
           for NC DIS at HERA, quark-pair production cross-section at LEP
           and Drell-Yan process at the Tevatron, as indicated in the plot.}
  \label{fig-diag}
\end{figure}

For leptoquark masses comparable with the available $ep$ center-of-mass energy
$u$-channel leptoquark exchange process and the $s$-channel leptoquark production
have to be considered separately. 
Corresponding diagrams for F=0 and F=2 leptoquarks are shown in
Figure \ref{fig-diag}.
The leptoquark contribution to the scattering amplitude 
can be now described by the following formulae:
\begin{itemize}
\item
for $u$-channel leptoquark exchange 
( $F$=0 leptoquark in $e^{-}q$ or $e^{+}\bar{q}$ scattering, 
  or $|F|$=2 leptoquark in $e^{+}q$ or $e^{-}\bar{q}$ scattering)
\begin{eqnarray}
\eta^{eq}_{ij}(s,u) & = & \frac{a^{eq}_{ij} \cdot \lambda_{LQ}^{2}}
                            { M_{LQ}^{2} - u }  \;\; , \nonumber
\end{eqnarray}

\item
for $s$-channel leptoquark production 
($F$=0 leptoquark in $e^{+}q$ or $e^{-}\bar{q}$ scattering, 
 or $|F|$=2 leptoquark in $e^{-}q$ or $e^{+}\bar{q}$ scattering)
\begin{eqnarray}
\eta^{eq}_{ij}(s,u) & = & \frac{a^{eq}_{ij} \cdot \lambda_{LQ}^{2}}
{M_{LQ}^{2} - s - i s \frac{\Gamma_{LQ}}{M_{LQ}}} \;\; , \nonumber
\end{eqnarray}
where $\Gamma_{LQ}$ is the total leptoquark width. 
The partial decay width for every decay channel is given by the formula:
\begin{eqnarray}
\Gamma_{LQ} & = & \frac{ \lambda_{LQ}^{2} M_{LQ} }
                       { 8 \pi ( J + 2 ) }  \;\; , \nonumber
\end{eqnarray}
where $J$ is the leptoquark spin.
\end{itemize}

For processes such as $e^{+}e^{-} \rightarrow hadrons$ a corresponding formula 
can be written for the $e^{+}e^{-} \rightarrow q \bar{q}$ 
tree level amplitude:
\begin{eqnarray}
M^{ee}_{ij}(s) & = & 
- \frac{4 \pi \alpha_{em} e_{q}}{s} \; + \;
  \frac{4 \pi \alpha_{em}}{\sin^{2}\theta_{W} \; \cos^{2}\theta_{W} } 
\cdot \frac{g^{e}_{i} g^{q}_{j}}
{s - M_{Z}^{2} + i s \frac{\Gamma_{Z}}{M_{Z}}}
  \; + \; \eta^{eq}_{ij}(t,u) \;\; , \label{eq-ms}
\end{eqnarray}
where the subscripts $i$ and $j$ label the chiralities of the initial 
lepton and final quark respectively and
\begin{eqnarray}
\eta^{eq}_{ij}(t,u) & = & 
\begin{array}\{{ll}.
\displaystyle
\frac{a^{eq}_{ij} \cdot \lambda_{LQ}^{2}}{ M_{LQ}^{2} - t }    &
{\rm for} \; F=0 \;\; ; \\[7mm] 
\displaystyle
\frac{a^{eq}_{ij} \cdot \lambda_{LQ}^{2}}{ M_{LQ}^{2} - u }    &
{\rm for} \; |F|=2 \;\; .\\
\end{array} \nonumber
%
%
\end{eqnarray}
Same formulae apply also to $q \bar{q} \rightarrow l^{+} l^{-}$ 
amplitude, with  $i$ and $j$ labelling the chiralities of the initial 
quark and final lepton respectively.

Leptoquark states with $\beta = \frac{1}{2}$ (coupling to both electron-quark 
and neutrino-quark pairs) contribute also to the charged current DIS at HERA
$e q \rightarrow \nu q'$.
For $M_{LQ} \gg \sqrt{s}$ the effective 
charged current contact interaction coupling is given by
\begin{eqnarray}
\eta^{CC} \equiv \eta^{eu \nu d} & = & 
             \left( a^{ed}_{LL} - a^{eu}_{LL} \right)
\cdot \left( \frac{\lambda_{LQ}}{M_{LQ}} \right)^{2}  \;\; . \label{eq-cc} 
\end{eqnarray}


\section{Experimental Data}
\label{sec-data}

\subsection{High-$Q^{2}$ DIS at HERA}
\label{sec-hera}

Used in this analysis are the 1994-97 data on high-$Q^{2}$ 
$e^{+}p$ NC DIS from both H1 \cite{h1pp} and ZEUS \cite{zeuspp}, 
as well as the recent results from $e^{-}p$ NC DIS 
scattering \cite{h1ep,zeusep}.
The analysis takes into account expected and 
measured numbers of events in bins of $Q^{2}$.
For simplicity let us consider a single $Q^{2}$ bin ranging from
$Q^{2}_{min}$ to $Q^{2}_{max}$. Assume that  $n_{SM}$ events are
expected from the Standard Model. 

The leading-order doubly-differential cross-section for positron-proton 
NC DIS ($e^{+}p \rightarrow e^{+} X$) can be written as 
\begin{eqnarray}
\frac{d^{2}\sigma^{LO}}{dx dQ^{2}} & = &
\frac{1}{16\pi} \sum_{q} 
q(x,Q^{2}) \left\{ |M^{eq}_{LR}|^{2} + |M^{eq}_{RL}|^{2} + 
(1-y)^{2} \left[  |M^{eq}_{LL}|^{2} + |M^{eq}_{RR}|^{2} \right] \right\}
\; +  \nonumber \\
 & &  ~~~~~~~~~~\bar{q}(x,Q^{2})\left\{ |M^{eq}_{LL}|^{2} + |M^{eq}_{RR}|^{2} + 
(1-y)^{2} \left[  |M^{eq}_{LR}|^{2} + |M^{eq}_{RL}|^{2} \right] \right\} \;\; ,
 \nonumber
\end{eqnarray}
where $x$ is the Bjorken variable, describing the fraction of the proton momentum 
carried by the struck quark (antiquark), $q(x,Q^{2})$ and $\bar{q}(x,Q^{2})$ are
the quark and antiquark momentum distribution functions in the proton and
$M^{eq}_{ij}$ are the scattering amplitudes of
equation \eqref{eq-mt}, which can include contributions from leptoquark
production or exchange processes.

The cross-section integrated over the $x$ and $Q^{2}$ range 
of an experimental $Q^{2}$ bin is
\begin{eqnarray}
\sigma^{LO} ( \lambda_{LQ} , M_{LQ}  ) & = & 
\bigint{Q^{2}_{min}}{Q^{2}_{max}} dQ^{2} 
\bigint{\frac{Q^{2}}{s \cdot y_{max}}}{1} dx \;  
    \frac{d^{2}\sigma^{LO}( \lambda_{LQ} , M_{LQ}  ) }{dx dQ^{2}} \label{eq-intdis}
\end{eqnarray}
where $y_{max}$ is an upper limit on the reconstructed Bjorken variable $y$,  
$y=\frac{Q^{2}}{x\;s}$,  imposed in the analysis.
The number of events expected from the Standard Model with 
leptoquark contributions can now be calculated as:
\begin{eqnarray}
n( \lambda_{LQ} , M_{LQ}  ) & = & n_{SM} \cdot 
\left( \frac{\sigma^{LO}( \lambda_{LQ} , M_{LQ}  ) }{\sigma^{LO}_{SM}} \right)  
                        \;\; ,          \label{eq-neta}
\end{eqnarray}
where $\sigma^{LO}_{SM}$ is the Standard Model cross-section
calculated with formula \eqref{eq-intdis} (setting $\lambda = 0$).
Leading-order expectations of the leptoquark models 
are used to rescale the Standard Model prediction $n_{SM}$ 
coming from detailed experiment simulation. 
This accounts for different experimental effects, and (to some extent) for
higher order QCD and electroweak corrections.\footnote{Correctly taken 
into account are only those corrections which are the same or similar 
for the Standard Model and for the cross-section including 
leptoquark contributions.} NLO QCD corrections to the resonant 
leptoquark production are introduced as an additional correction 
factor, based on \cite{kqcd}.

For models with leptoquarks coupling to both electron-quark and 
neutrino-quark pairs ($S_0^L$, $S_1$, $V_0^L$ and $V_1$),
HERA data on $e^{+}p$ and  $e^{-}p$ CC DIS \cite{h1pp,h1ep,zeuscc} 
are also included in the fit.

\begin{figure}[tbp]
\centerline{\resizebox{\figwidth}{!}{%
  \includegraphics{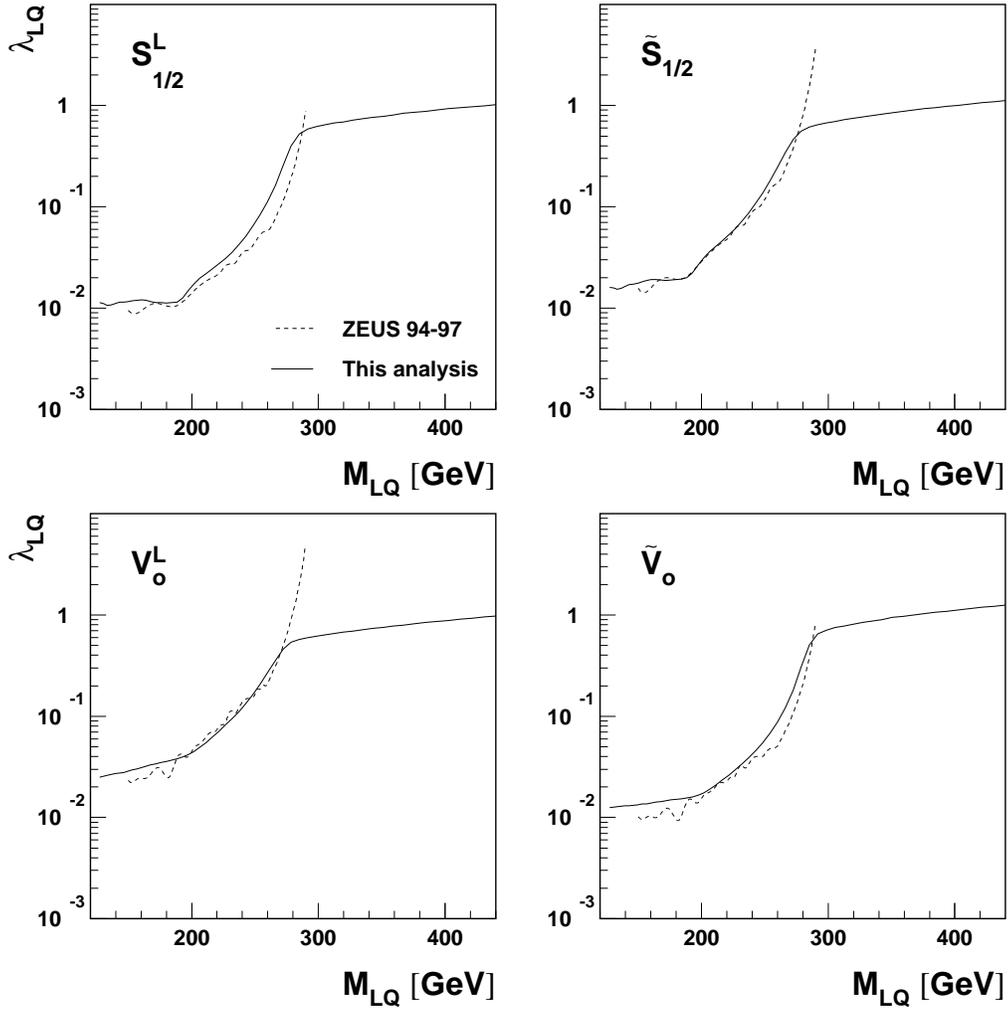}
}}
  \caption{Comparison between limits calculated from 
   the $Q^{2}$ distribution of the ZEUS $e^{+}p$ NC DIS data 
   (this analysis)  and the published ZEUS limits \cite{zeuslq}
   for selected F=0 leptoquarks, as indicated in the plot.  }
  \label{fig-lqres5}
\end{figure}

In the limit of heavy leptoquark masses ($M_{LQ} \gg \sqrt{s}$) 
the $Q^{2}$ distribution of NC and CC DIS events is
most sensitive  to the leptoquark couplings. 
For masses below $\sqrt{s} \sim $300 GeV, where direct leptoquark
production becomes  possible at HERA, better limits are obtained from
studying the electron-jet invariant mass distribution.
However, to describe correctly the narrow leptoquark resonance
production and reconstruction, sizable QED and QCD corrections
as well as complicated detector effects have to be taken into account.
As these corrections could not be included in the analysis,
the $Q^{2}$ distribution was used to constrain leptoquark couplings
in the whole mass range. 
Comparison between limits calculated from the $Q^{2}$ distribution
of the ZEUS $e^{+}p$ NC DIS data \cite{zeuspp} and the published
ZEUS limits for F=0 leptoquarks \cite{zeuslq}\footnote{
Similar limits on the leptoquark couplings and masses 
have also been presented by the H1 Collaboration \cite{h1lq}.}
is presented in Figure \ref{fig-lqres5}.
Taking into account that ZEUS analysis includes mass dependent 
selection cuts and that it was optimised for leptoquark search,
the difference between the two approaches is surprisingly small.
Direct ZEUS limits are at most 40\% lower (depending on the 
model and the mass range) than the one obtained from 
the $Q^{2}$ distribution. 

\subsection{Measurements from LEP}
\label{sec-lep}

Many measurements at LEP  are sensitive to different kinds of "new physics".  
The leptoquark exchange contribution can be directly tested in the
measurement of the total hadronic cross-section 
above the $Z^{\circ}$ pole.\footnote{For the leptoquark masses and couplings
considered here the effects of the possible leptoquark exchange at 
$\sqrt{s}=M_{Z}$ are completely negligible in comparison with the resonant
$Z^{\circ}$ production.}
The leading order formula for the total quark pair production
cross-section, $\sigma (e^{+} e^{-} \rightarrow q \bar{q})$, at an
electron-positron center-of-mass energy squared, $s$, is
\begin{eqnarray}
\sigma^{LO}(s) & = &
\frac{3 s}{128\pi} \sum_{q} 
\; \int d \cos \theta  \;   \nonumber \\
& &
\left[
\left( |M^{ee}_{LL}|^{2} + |M^{ee}_{RR}|^{2} \right) (1+ \cos \theta)^2 +
\left( |M^{ee}_{LR}|^{2} + |M^{ee}_{RL}|^{2} \right) (1- \cos \theta)^2
\right] \;\; ,
\label{eq-lep}
\end{eqnarray}
where $M^{ee}_{ij}$ are 
the scattering amplitudes described by equation \eqref{eq-ms}, 
including contributions from leptoquark exchange and
$\theta$ is the quark production angle in the $e^{+} e^{-}$
center-of-mass system.
For comparison with measured experimental values, the expected 
Standard Model cross-section $\sigma^{SM}(s)$ quoted by experiments
are rescaled using the ratio of the leading order
cross-sections with and without leptoquark contribution:
\begin{eqnarray}
\sigma(s, \lambda_{LQ}, M_{LQ} ) & = &
\sigma^{SM}(s) \cdot \left(
\frac{\sigma^{LO}(s,\lambda_{LQ}, M_{LQ} )}
     {\sigma^{LO}_{SM}} \right) \;\; , \label{eq-seta}
\end{eqnarray}
where $\sigma^{LO}_{SM}$ is the leading-order Standard Model 
cross-section ($\lambda=0$), calculated with
equation \eqref{eq-lep}.
This takes into account possible experimental effects and
higher order QCD and electroweak corrections.
Included in the analysis are data on $\sigma_{had}$ 
from Aleph, Delphi, L3 and Opal experiments for center-of-mass
energies up to 202~GeV \cite{aleph,delphi,l3,opal}.
All measurements are in good agreement with the Standard Model 
predictions.
However, cross-section values obtained for $\sqrt{s}$=192--202 GeV
are on average about 2.5\% above the predictions.
The combined significance of this deviation is only about 
2.3$\sigma$ \cite{lepnew} but it has an important influence 
on the global analysis.

In the global analysis of electron-quark contact 
interactions \cite{mygcia}, the strongest constraints on
the contact interaction couplings resulted from the LEP data
on heavy quark production, $R_{q}$ ($q=b,c$),
and on forward-backward asymmetries, $A^{q}_{FB}$.
However, this is only the case for models assuming family universality.
For the first-generation leptoquarks, the constraints resulting from
LEP measurements based on heavy flavour tagging are much weaker
than those resulting from hadronic cross-section measurements.
Nevertheless, possible deviations in the $u\bar{u}$ and $d\bar{d}$ 
quark pair production cross-sections (resulting in the deviation
of the total hadronic cross-section) can be also constrained 
using results on $R_{c}$ and $R_{b}$.
Results on $A^{q}_{FB}$ are included in the presented analysis for consistency with
the previous study \cite{mygcia}.

\subsection{Drell-Yan lepton pair production at the Tevatron}
\label{sec-dy}

Used in this analysis are data on Drell-Yan electron pair 
production ($p \bar{p} \rightarrow e^{+} e^{-}\;X$)
from the CDF \cite{dy_cdf} and D$\emptyset$ \cite{dy_d0} experiments.
The leading order cross-section for lepton pair production in 
$p \bar{p}$ collisions is
\begin{eqnarray}
\frac{d^{2}\sigma^{LO}}{dM_{ll} dY} & = &
\frac{M_{ll}^{3}}{192\pi s} \sum_{q} q(x_{1}) q(x_{2}) \cdot  \nonumber \\
 & &  \int  d\cos \theta \left[
\left( |M^{ee}_{LL}|^{2} +  |M^{ee}_{RR}|^{2} \right) (1+\cos \theta)^2 +
\left( |M^{ee}_{LR}|^{2} +  |M^{ee}_{RL}|^{2} \right) (1-\cos \theta)^2
\right] \;\; ,
\nonumber
\end{eqnarray}
where $M_{ll}$ is the invariant lepton pair mass, $Y$ is the rapidity 
of the lepton pair, $\theta$ is the lepton
production angle in their center-of-mass system, and $x_{1}$ and $x_{2}$ are
the fractions of the proton and antiproton momenta carried by the annihilating
$q \bar{q}$. 
When integrating over $\theta$, the angular detector coverage is taken 
into account.
The scattering amplitudes 
$M^{ee}_{i j}$ and the parton density functions are calculated at
a mass scale
\begin{eqnarray}
\mu^2 \; = \; = \hat{s} & = & x_{1} x_{2} s  \;\; , \nonumber
\end{eqnarray}
where $s$ is the total proton-antiproton center-of-mass energy squared.

The cross-section corresponding to the $M_{ll}$ range from $M_{min}$ to
$M_{max}$ is calculated as
\begin{eqnarray}
\sigma^{LO} ( \lambda_{LQ}, M_{LQ} ) & = & 
\bigint{M_{min}}{M_{max}} dM_{ll} 
\bigint{-Y_{max}}{Y_{max}} dY  \; 
  \frac{d^{2}\sigma ( \lambda_{LQ}, M_{LQ}) }{dM_{ll} dY} \;\; ,
\label{eq-intdy}
\end{eqnarray}
where $Y_{max}$ is the upper limit on the rapidity of the produced 
lepton pair:
\begin{eqnarray}
Y_{max} & = & \ln \frac{\sqrt{s}}{M_{ll}} \;\; . \nonumber 
\end{eqnarray}
The cross-section calculated with equation \eqref{eq-intdy} is used to 
calculate the number of events expected from the Standard Model with 
leptoquark contribution using formula \eqref{eq-neta}.

\subsection{Direct limits from the Tevatron}
\label{sec-tevdir}

The D$\emptyset$ and CDF experiments at the Tevatron presented limits 
on the first-generation scalar leptoquark masses from the search
for leptoquark pair production in hard interactions
($p \bar{p} \rightarrow LQ \; \overline{LQ} \; X$).
Both experiments see no leptoquark candidate events,
with leptoquarks decaying into an electron and a jet, above 
a reconstructed leptoquark mass of 200 GeV \cite{d0lq,cdflq}.
The result of the NLO cross-section 
calculations\footnote{assuming mass scale $\mu = 2 M_{LQ}$} 
\cite{tevnlo} can be parameterised in this mass region as
\begin{eqnarray}
\sigma^{S}_{LQ}(M_{LQ}) & \approx & 
 114.6 \; {\rm pb} \; \cdot \; 
   \exp \left( -\frac{M_{LQ}}{30.28 \; {\rm GeV}} \right) \;\; . \nonumber
\end{eqnarray}
The expected number of leptoquark events reconstructed 
in $eejj$ channel is
\begin{eqnarray}
n_{exp}(M_{LQ}) & = & 
       \epsilon  {\cal L} \cdot \sigma^{S}_{LQ}(M_{LQ}) 
                 \; \sum_{LQ} \beta_{LQ}^{2} \;\; , \nonumber
\end{eqnarray}
where the sum is over leptoquark states within 
the considered multiplet and the combined effective luminosity 
(i.e. luminosity corrected for selection efficiency) 
for two experiments  is
$ \epsilon  {\cal L} \approx 78 \; {\rm pb}^{-1}$.
For leptoquark states with $\beta = 0.5$, the results
of D$\emptyset$ search in $e \nu j j $  channel are also included in
the analysis.
Because of the assumed mass degeneration
the mass limits for scalar leptoquark multiplets
can be significantly higher than for single leptoquarks.
For $S^{R}_{1/2}$ isospin doublet ($\sum \beta_{LQ}^{2}$=2) the combined
limit is $M_{LQ} >$ 263 GeV, as compared with
the published limit of 242 GeV for single leptoquark 
production \cite{cdfd0}.

For vector leptoquarks, pair production cross-section
at the Tevatron strongly depends on the unknown (not constrained in
the BRW model) anomalous leptoquark couplings.
For the presented analysis the values giving the smallest
vector leptoquark pair production cross-section  were
assumed \cite{lqmin}.
The LO vector leptoquark production cross-section has been parametrised as:
\begin{eqnarray}
\sigma^{V}_{LQ}(M_{LQ}) & \approx & 
 268.1 \; {\rm pb} \; \cdot \; 
   \exp \left( -\frac{M_{LQ}}{28.65 \; {\rm GeV}} \right) \;\; . \nonumber
\end{eqnarray}
Only the D$\emptyset$ experiment has presented limits on the vector leptoquark
masses in the minimum cross-section model \cite{d0vec}. 
The limit  $M_{LQ}>$ 245~GeV (for $\beta$=1) corresponds to the effective
luminosity of $ \epsilon  {\cal L} \approx 58 \; {\rm pb}^{-1}$.

\subsection{Data from low energy experiments}
\label{sec-le}

The low energy data are included in the present analysis in 
exactly the same way as in the contact interaction analysis \cite{mygcia}.
For all leptoquark models the following
constraints from low energy experiments are considered:
\begin{itemize}
\item Atomic Parity Violation (APV) \\
The Standard Model predicts parity non-conservation in atoms caused
(in lowest order) by the $Z^{\circ}$ exchange between
electrons and quarks in the nucleus. 
Experimental results on parity violation in atoms are given in terms
of the weak charge $Q_{W}$ of the nuclei. 
Standard Model prediction for $Q_{W}$ are based on the very
precise measurement of the $\sin^{2} \Theta_{W}$ at LEP1 and SLD.
A new determination
of $Q_{W}$ for Cesium atoms was recently reported \cite{apvnew}.
The experimental result differs from the Standard Model 
prediction by:
\begin{eqnarray}
\Delta Q_{W}^{Cs}  \equiv Q_{W}^{meas} - Q_{W}^{SM} 
                      & = & 1.13 \pm 0.46   \nonumber
\end{eqnarray}
As already mentioned in the Introduction, this 2.5$\sigma$ discrepancy
between the measurement and Standard Model predictions 
induces significant evidence for some leptoquark models. 
Also other "new physics" processes,
as for example $Z^{\circ \prime}$ exchange,
were proposed as a possible explanation 
for the APV measurement.
One has to take into account that these new processes 
can also affect precision measurements
at LEP1 and the determination of  $\sin^{2} \Theta_{W}$,
making the analysis much more difficult.
However, for the leptoquark masses and couplings
considered here the effects 
of the possible leptoquark exchange at $\sqrt{s}=M_{Z}$ 
can be safely neglected.

The leptoquark
contributions to $Q_{W}$ is:
\begin{eqnarray}
\Delta Q_{W} & = & 
\frac{2Z+N}{\sqrt{2} G_{F}}
 \left(\eta^{eu}_{LL}+\eta^{eu}_{LR}-\eta^{eu}_{RL}-\eta^{eu}_{RR} \right)
                                   \nonumber \\
 & + &
\frac{Z+2N}{\sqrt{2} G_{F}}
 \left(\eta^{ed}_{LL}+\eta^{ed}_{LR}-\eta^{ed}_{RL}-\eta^{ed}_{RR} \right)
\;\; ,       \nonumber
\end{eqnarray}
where $\eta^{eq}_{ij}$ are the effective couplings given 
by \eqref{eq-cieta}.

\item Electron-nucleus scattering \\
The limits on possible leptoquark contributions to 
electron-nucleus scattering at low energies can be extracted from
the polarisation asymmetry measurement
\begin{eqnarray}
A  & = & 
\frac{d\sigma_{R}-d\sigma_{L}}{d\sigma_{R}+d\sigma_{L}} \;\; , \nonumber
\end{eqnarray}
where $d\sigma_{L(R)}$ denotes the differential cross-section of
left- (right-) handed electron scattering. 
Polarisation asymmetry directly measures the parity violation
resulting from the interference between $Z^{\circ}$ and $\gamma$
scattering amplitudes.
For isoscalar targets, taking into account valence quark contributions only,
the polarisation asymmetry for elastic electron scattering is
\begin{eqnarray}
A  & = & 
-\frac{3 \sqrt{2} G_{F} Q^{2}}{20 \pi \alpha_{em} } 
\left[ 2 \left( g^{u}_{L} + g^{u}_{R} \right) - 
         \left( g^{d}_{L} + g^{d}_{R} \right) \right]  \;\; , \nonumber
\end{eqnarray}
where $Q^{2}$ is the four-momentum transfer and the effective 
electroweak coupling of the quark is modified by the leptoquark 
contribution
\begin{eqnarray}
 \bigmod{g^{q}_{i}}{ef\!f} & = & 
  g^{q}_{i} - \frac{\eta^{e q}_{Li}}{2\sqrt{2} G_{F}} \;\; . \label{eq-geff}
\end{eqnarray}
The data used in this analysis come from the SLAC $e$D 
experiment \cite{slac}, the Bates $e$C experiment \cite{bates} and 
the Mainz experiment on $e$Be scattering \cite{mainz}. 

\end{itemize}

For leptoquarks contributing to charged current processes, 
additional constraints come from:
\begin{itemize}

\item Lepton-hadron universality of weak Charged Currents \\
New charged current interactions would  affect the
measurement of $V_{ud}$ element of the Cabibbo-Kobayashi-Maskawa (CKM) 
matrix, leading to an effective violation of unitarity
\cite{cickm,cickm2}. The new experimental constraint is \cite{rpp}
\begin{eqnarray}
|V_{ud}|^{2} + |V_{us}|^{2} + |V_{ub}|^{2} & = & 0.9959 \pm 0.0019 
\;\; , \nonumber
\end{eqnarray}
whereas the expected leptoquark contribution is
\begin{eqnarray}
V_{ud} & = & 
V_{ud}^{SM} \cdot \left( 1 - \frac{\eta^{CC}}{2\sqrt{2}G_{F}} \;\; , \right) 
\nonumber
\end{eqnarray}
with $\eta^{CC}$ is given by equation \eqref{eq-cc}.

\item Electron-muon universality \\
In the similar way new charged current interactions
would also lead to effective violation of $e$-$\mu$ universality
in charged pion decay \cite{cickm}. The current experimental value of
$R=\Gamma(\pi^{-} \rightarrow e \bar{\nu})/
   \Gamma(\pi^{-} \rightarrow \mu \bar{\nu})$ is \cite{emu}
\begin{eqnarray}
\frac{R_{meas}}{R_{SM}} & = & 0.9966 \pm 0.030 \;\; , \nonumber
\end{eqnarray}
whereas the expected contribution from leptoquark exchange is
\begin{eqnarray}
\bigmod{R}{meas} & = & 
R_{SM} \cdot \left( 1 - \frac{\eta^{CC}}{2\sqrt{2}G_{F}} \right)^{2} 
\;\; . \nonumber
\end{eqnarray}

\end{itemize}

It is important to notice, that data in the charged current sector 
also indicate a possible deviation from the Standard Model predictions:
a slight violation of the unitarity of the CKM matrix 
and of the $e$-$\mu$ universality. 
The combined significance of these two results is about 2.4$\sigma$
and has a considerable influence on the presented analysis.


\section{Analysis method}
\label{sec-method}

\rightmark

The analysis method is similar to the one used in 
the recently published analysis \cite{mygcia}.
For every leptoquark coupling and mass value the probability function
describing the agreement between the model and the data is calculated:
\begin{eqnarray}
{\cal P}(\lambda_{LQ},M_{LQ}) & \sim & 
\prod_{i} P_{i}(\lambda_{LQ},M_{LQ}). \label{eq-prob}
\end{eqnarray}
The product runs over all experimental 
data $i$. 
The logarithm of the probability function $\ln {\cal P}$
is the so called log-likelihood function,
which is often used in similar analysis:
\begin{eqnarray}
\ln {\cal P}(\lambda_{LQ},M_{LQ}) & = & 
\sum_{i} \ln P_{i}(\lambda_{LQ},M_{LQ}). \nonumber
\end{eqnarray}
The data used in this analysis can be divided 
into two classes.
\begin{enumerate}
\item
For experiments in which a result is presented as 
a single number with an error which is considered to reflect a Gaussian
probability distribution, the probability function 
can be written as
\begin{eqnarray}
P_{i}(\lambda_{LQ},M_{LQ}) & \sim & 
exp \left( -\frac{1}{2} 
\frac{(F(\lambda_{LQ},M_{LQ}) - \Delta A)^{2}}{\sigma_{A}^{2}} \right)
\;\; , \label{eq-gauss}
\end{eqnarray}
where $\Delta A$ is the difference between the measured value and the Standard
Model prediction,  $\sigma_{A}$ is the measurement error and 
$F(\lambda_{LQ},M_{LQ})$ is the expected leptoquark contribution 
to the measured value.
This approach is used for all low energy data as well as for
the LEP hadronic cross-section measurements.

\item
On the other hand, if the experimentally measured quantity is 
the number of events of a particular kind (e.g. HERA high-$Q^{2}$ data
or Drell-Yan lepton pairs and direct search results from the Tevatron), 
and especially when this number is small, the probability 
is better described by the Poisson distribution
\begin{eqnarray}
P_{i}(\lambda_{LQ},M_{LQ}) & \sim & 
\frac{ n(\lambda_{LQ},M_{LQ})^{N} \cdot exp( -n(\lambda_{LQ},M_{LQ})) }
     { N ! } \;\; ,
\label{eq-poisson}
\end{eqnarray}
where $N$ and $n(\lambda_{LQ},M_{LQ})$ are the measured and expected 
number of events in a given experiment, respectively, and 
$n(\lambda_{LQ},M_{LQ})$ takes into account a possible
leptoquark contribution. This approach has been used for
the HERA and the Tevatron data.

\end{enumerate}

For low energy data the total measurement error can be used in 
\eqref{eq-gauss} taking into account both statistical and
systematic errors.
For collider data, formula \eqref{eq-gauss} or \eqref{eq-poisson} 
is used to take into account the statistical error of the measurement only. 
The systematic errors are assumed to be correlated to 100\%
within a given data set (e.g. $e^{+}p$ NC DIS data from ZEUS ) they
This approach, as well as the migration corrections used for
HERA and Tevatron Drell-Yan results are discussed in detail 
in \cite{mygcia}.

The probability function ${\cal P}(\lambda_{LQ},M_{LQ})$ 
summarises our current experimental knowledge about possible 
leptoquark couplings and masses.
As ${\cal P}$ is not a probability distribution, it does
not satisfy any normalisation condition.
Instead it is convenient to rescale the probability function in such a
way that for the Standard Model it has the value of 1:
\begin{eqnarray}
{\cal P}(\lambda_{LQ}=0,M_{LQ})  & = &  1.  \label{eq-pmax}  \\
{\rm or} \;\;\;\;  
\ln {\cal P}(\lambda_{LQ}=0,M_{LQ})  & = &  0. \nonumber
\end{eqnarray}

Using the probability function ${\cal P}(\lambda_{LQ},M_{LQ})$ 
two types of limits in $(\lambda_{LQ},M_{LQ})$  space are calculated:
\begin{itemize}

\item
      Rejected are all models (parameter values) which result in
%
\begin{eqnarray}
{\cal P}(\lambda_{LQ},M_{LQ})  & < &  0.05 \nonumber \\
{\rm or} \;\;\;\;  
\ln {\cal P}(\lambda_{LQ},M_{LQ}) & < &  -3.0 \nonumber 
\end{eqnarray}
This is taken as the definition of the 95\% confidence level (CL)
{\bf exclusion limit}. Exclusion limits presented in this paper 
are lower limits in case of leptoquark mass $M_{LQ}$ and
upper limits in case of $\lambda_{LQ}$ or $\lambda_{LQ} / M_{LQ}$.

\item
      Some leptoquark models turn out to describe the data
      much better than the Standard Model:
\begin{eqnarray}
{\cal P}_{max} \; \equiv \;
\max_{\lambda_{LQ},M_{LQ}} {\cal P}(\lambda_{LQ},M_{LQ}) 
                    & \gg &  1.  \nonumber
\end{eqnarray}
In that case the 95\% CL {\bf signal limit}
corresponding to the uncertainty on the "best" values of
 $\lambda_{LQ}$ and $M_{LQ}$ is defined by
the condition
\begin{eqnarray}
{\cal P}(\lambda_{LQ},M_{LQ})  & > &  0.05 \cdot  {\cal P}_{max} \nonumber \\
{\rm or} \;\;\;\;  
\ln {\cal P}(\lambda_{LQ},M_{LQ}) & < &  \ln {\cal P}_{max} \; - \; 3.0 \nonumber
\end{eqnarray}
\end{itemize}

In the previous analysis \cite{mygcia} no significant 
deviations from the Standard Model were observed. 
In such a case both definitions give similar results
and there is no need to distinguish between exclusion 
and signal limits.


\section{Results}
\label{sec-results}

\begin{figure}[tbp]
\centerline{\resizebox{\figwidth}{!}{%
  \includegraphics{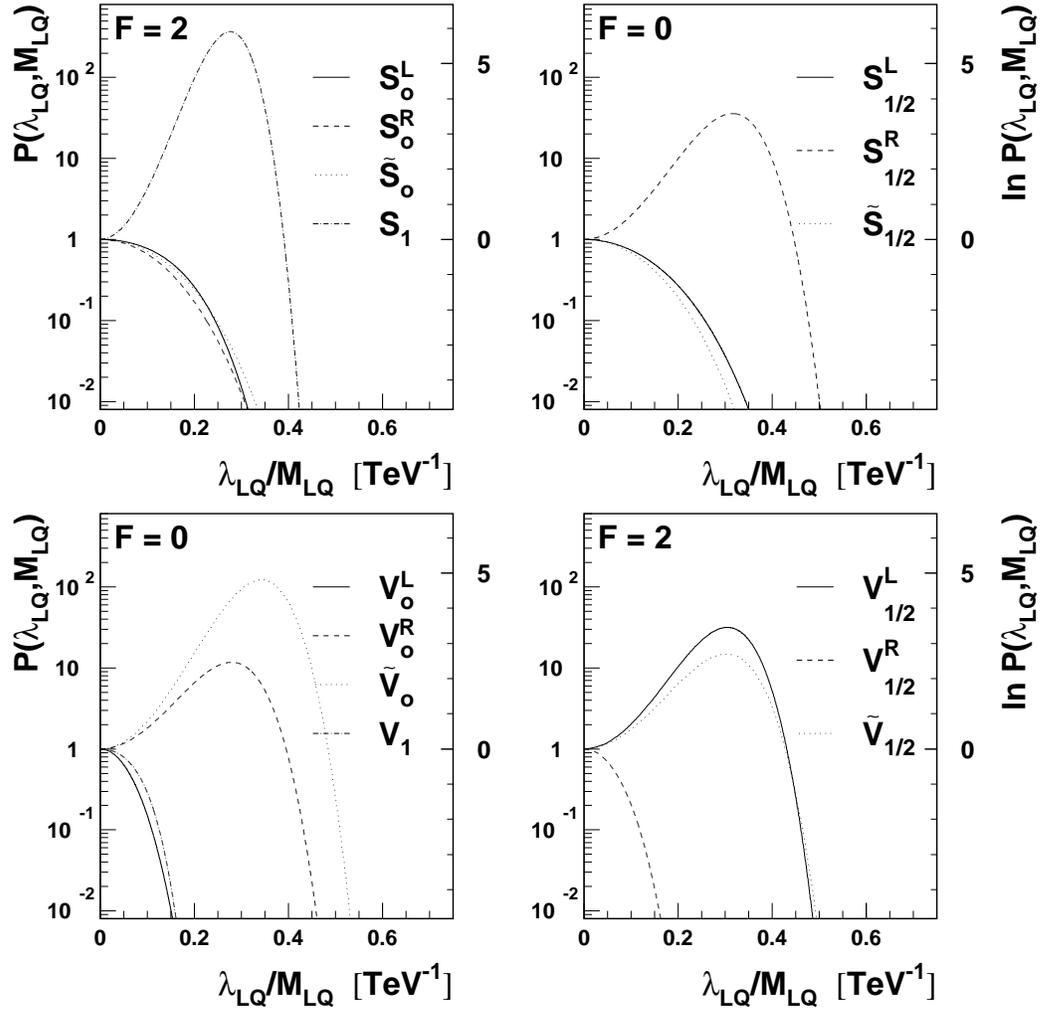}
}}
  \caption{ Probability function ${\cal P}(\lambda_{LQ},M_{LQ})$ (left hand scale)
         and the log-likelihood function  ${\ln \cal P}(\lambda_{LQ},M_{LQ})$
           (right hand scale), 
            in the limit of very high leptoquark masses, 
            for different leptoquark models  as indicated on the plot.}
  \label{fig-lqres1}
\end{figure}

In the limit of very high leptoquark masses (contact interaction
approximation) the probability function depends only on the
$\lambda_{LQ}/M_{LQ}$ ratio.
Using the global model probability ${\cal P}(\lambda_{LQ},M_{LQ})$,
as defined by equation \eqref{eq-prob}, 
the value $\left( \lambda_{LQ}/M_{LQ} \right)_{max}$  
giving the maximum probability is determined for each model.
The results are presented in Table \ref{tab-lqres1}.
The attributed errors, quoted for models which give better description
of the data than the Standard Model 
(i.e.  $\left( \lambda_{LQ}/M_{LQ} \right)_{max} > 0$)
correspond to the decrease in $\ln {\cal P}(\lambda_{LQ},M_{LQ})$ 
by $\frac{1}{2}$.
\begin{table}[tbp]
  \begin{center}
    \begin{tabular}{llrrccc}
      \hline\hline\hline\noalign{\smallskip}
    & \multicolumn{3}{c}{best description} 
    & 95\% CL signal  & \multicolumn{2}{c}{95\% CL exclusion limits } \\    
    \cline{2-4} \cline{6-7} \noalign{\smallskip}
Model & $\left(\frac{\lambda_{LQ}}{M_{LQ}}\right)_{max}$ 
          & ${\cal P}_{max}$     & $\ln {\cal P}_{max}$  
          &  $\left(\frac{\lambda_{LQ}}{M_{LQ}}\right)$   
          &  $\left(\frac{\lambda_{LQ}}{M_{LQ}}\right)$   & $M_{LQ}$  \\
     & ${\rm [TeV^{-1}]}$  &  &  &   [TeV$^{-1}$]    & [TeV$^{-1}$] &  [GeV]\\ 
\hline\hline\hline\noalign{\smallskip}
%
%
$S_{\circ}^L$  &  0        &  1.0   &  0.0  &  
              & 0.27  &   213  \\ 
$S_{\circ}^R$  &  0        &  1.0   &  0.0  &  
              & 0.25   &   242  \\ 
$\tilde{S}_{\circ}$  &  0        &  1.0   &  0.0  &  
              & 0.28   &   242  \\ 
$S_{1/2}^L$  &  0        &  1.0   &  0.0  &  
              & 0.29  &   229  \\ 
$S_{1/2}^R$  & 0.32 $\pm$ 0.06 &  35.8 & 3.6 & 
    0.09--0.44 &  0.49  &   245  \\  
$\tilde{S}_{1/2}$  &  0        &  1.0   &  0.0  &  
              & 0.26   &   233  \\ 
$S_1$  & 0.28 $\pm$ 0.04 &  367. & 5.9 & 
    0.15--0.36 &  0.41  &   245  \\  
\hline\noalign{\smallskip}
$V_{\circ}^L$  &  0        &  1.0   &  0.0  &  
              & 0.12  &   230  \\ 
$V_{\circ}^R$  & 0.28 $\pm$ 0.07 &  11.7 & 2.5 & 
              & 0.44   &   231  \\ 
$\tilde{V}_{\circ}$  & 0.34 $\pm$ 0.06 &  122. & 4.8 & 
    0.16--0.46 &  0.52  &   235  \\  
$V_{1/2}^L$  & 0.30 $\pm$ 0.06 &  31.7 & 3.5 & 
    0.08--0.42 &  0.47   &   235  \\  
$V_{1/2}^R$  &  0        &  1.0   &  0.0  &  
              & 0.13   &   262  \\ 
$\tilde{V}_{1/2}$  & 0.30 $\pm$ 0.07 &  14.8 & 2.7 & 
              & 0.47   &   244  \\ 
$V_1$  &  0        &  1.0   &  0.0  &  
              & 0.14   &   254  \\ 
\hline\hline\hline\noalign{\smallskip}
    \end{tabular}
  \end{center}
  \caption{Coupling to mass ratio,
    $\left(\frac{\lambda_{LQ}}{M_{LQ}}\right)_{max}$,
 resulting in the best description  of the experimental data
 in the contact interaction approximation,   
 and the corresponding model probability ${\cal P}_{max}$  
and the log-likelihood ${ln \cal P}_{max}$, 
for  different leptoquark models, as indicated in the table. 
The errors attributed to non-zero $\frac{\lambda_{LQ}}{M_{LQ}}$ values 
correspond to the decrease of $\ln {\cal P}$ by $\frac{1}{2}$. 
Also given are 95\% CL signal  (for models with ${\cal P}_{max}>20$)  
and (upper) exclusion limits on  $\frac{\lambda_{LQ}}{M_{LQ}}$, and
(lower) exclusion limits on leptoquark masses $M_{LQ}$.}
  \label{tab-lqres1}
\end{table}
The probability functions ${\cal P}(\lambda_{LQ},M_{LQ})$ for different 
leptoquark models are shown in Figure \ref{fig-lqres1}.

For 8 out of 14 leptoquark models, the Standard Model gives the best 
description of the considered experimental data 
($\left( \lambda_{LQ}/M_{LQ} \right)_{max} = 0$). 
95\% CL exclusion limits for $\lambda_{LQ}/M_{LQ}$
range for these models  from 0.12 TeV$^{-1}$ 
(for $V_{\circ}^{L}$ model) to 0.29  TeV$^{-1}$ (for $S^{L}_{1/2}$ model).
The other 6 models are able to describe the data
better than the Standard Model.
In all cases the "best" coupling to mass ratio turns out to
be of the order of 0.3~TeV$^{-1}$.

\begin{figure}[tbp]
\centerline{\resizebox{\figwidth}{!}{%
  \includegraphics{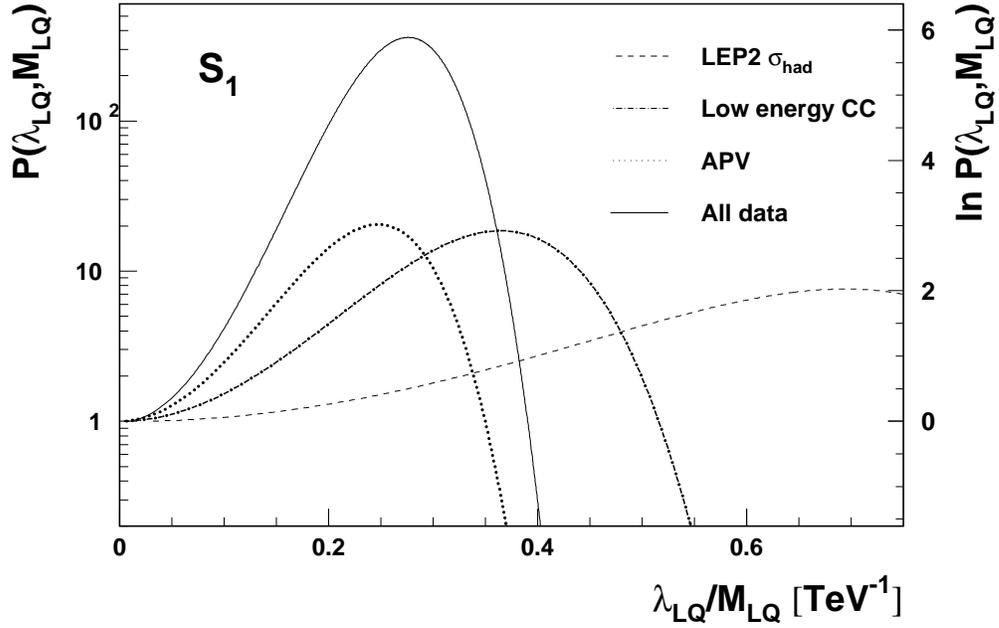}
}}
  \caption{ Contributions of different data sets 
            (as indicated on the plot) to the global
            probability functions ${\cal P}(\lambda_{LQ},M_{LQ})$ (left hand scale)
    and to the log-likelihood function  ${\ln \cal P}(\lambda_{LQ},M_{LQ})$
           (right hand scale), 
            for the $S_{1}$ model in the limit of very high 
            leptoquark masses.}
  \label{fig-lqres6a}
\end{figure}

The best description of the data is given by the  $S_{1}$ model
for $\left( \lambda_{LQ}/M_{LQ} \right)_{max} = 0.28 \pm 0.04$ TeV$^{-1}$
resulting in the maximum probability ${\cal P}_{max}$=367 
($\ln {\cal P}_{max}$=5.9). For the Gaussian probability function
this would correspond to about 3.4$\sigma$ deviation from the Standard Model.
The effect is mainly due to the APV result: 
the contribution of the APV measurement to the maximum
probability is $P$=20 ($\ln P$=3.0), 
corresponding to a 2.4$\sigma$ deviation from the Standard Model.
The result is also supported by the low energy charged current
data (unitarity of the CKM matrix and $e$-$\mu$ universality; 
$\ln P$=2.4,  2.2$\sigma$ effect) and LEP2 hadronic cross-section 
measurements ($\ln P$=0.5, 1.0$\sigma$ effect).
Contributions of different
data sets to the $S_{1}$ model probability function
are presented in Figure \ref{fig-lqres6a}. 
The fitted value of $\left( \lambda_{LQ}/M_{LQ} \right)_{max}$
results in almost the best description of both APV and low energy
charged current data, whereas LEP2 hadronic cross-section measurements
suggest even higher values of 
$\lambda_{LQ}/M_{LQ} \sim 0.7$ TeV$^{-1}$.
The 95\% CL signal limit, corresponding to model the probabilities
${\cal P} > 0.05\cdot {\cal P}_{max}$ is
$ 0.15 < \lambda_{LQ}/M_{LQ} < $0.36~TeV$^{-1}$.

The $\tilde{V_{\circ}}$ model also gives a very good description of
the data, resulting in ${\cal P}_{max}$=122
($\ln P$=4.8 corresponding to about 3.1$\sigma$).  
In this case the APV result ($\ln P$=3.1, 2.5$\sigma$) 
is strongly supported by LEP2 data ($\ln P$=1.3, 1.6$\sigma$).
The $S_{1/2}^{R}$ and $V_{1/2}^{L}$ models describe 
the APV measurement as well but they do not improve the description of
other data. 
For $V_{\circ}^{R}$ and $\tilde{V}_{1/2}$ models, the coupling values
required to explain APV data are disfavoured by other experiments
(mainly by LEP2 hadronic cross-section measurements) resulting in
even smaller ${\cal P}_{max}$ values.
Signal limits for 4 models which result in ${\cal P}_{max} > 20 $
are included in Table \ref{tab-lqres1}.
For models with  ${\cal P}_{max}>1$ (models describing the APV data) 
the 95\% CL exclusion limits on  $\lambda_{LQ}/M_{LQ}$
range from 0.41 TeV$^{-1}$ 
(for $S_{1}$ model) to 0.52  TeV$^{-1}$ (for $\tilde{V}_{\circ}$ model).

All of the results presented above were based on 
the contact interaction approximation, which is valid
for the leptoquark masses above about 1 TeV.
In the second part of the presented analysis 
lower leptoquark masses were also considered.
In that case, leptoquark constraints have to be studied in terms of
the leptoquark coupling and the leptoquark mass 
as two independent parameters.

\begin{figure}[tbp]
\centerline{\resizebox{\figwidth}{!}{%
  \includegraphics{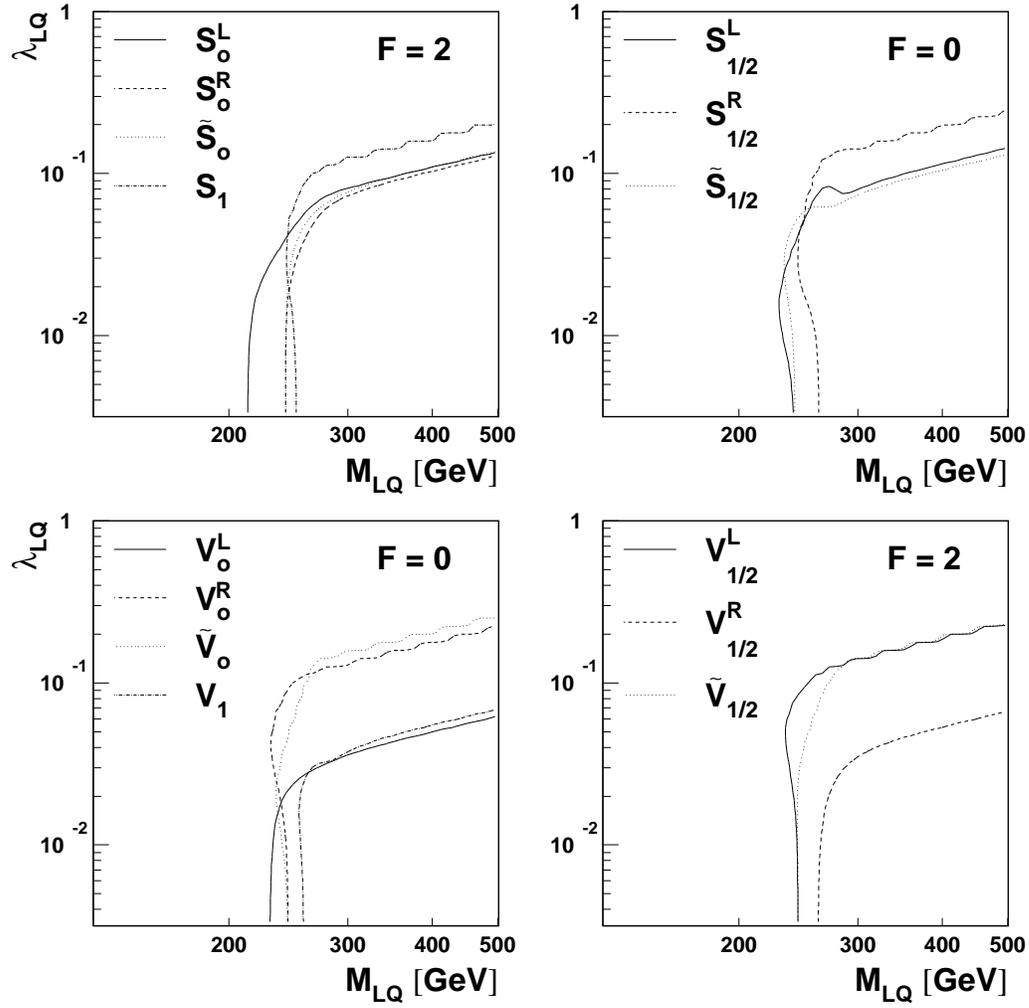}
}}
  \caption{ 95\% CL exclusion limits in  $(\lambda_{LQ},M_{LQ})$
            space, for different leptoquark 
            models, as indicated in the plot.
            Excluded are coupling and mass values above or
            to the left of the limit curves.}
  \label{fig-lqres2}
\end{figure}

Below 1 TeV, effects of the finite leptoquark mass reduce 
the virtual leptoquark exchange contribution to the expected LEP and
Tevatron cross-sections.
However, this effect is small and the asymptotic limit on
$\lambda_{LQ}/M_{LQ}$ increases by only about 10\% for leptoquark
masses $M_{LQ} \sim $ 300 GeV.
For masses below 300 GeV, the limits on $\lambda_{LQ}$ become much stronger
because of the direct searches at HERA and at the Tevatron.
Combined constraints on the leptoquark coupling and mass are
derived from the probability function ${\cal P}(\lambda_{LQ},M_{LQ})$,
as described in section \ref{sec-method}.
The 95\% CL exclusion limits in $(\lambda_{LQ},M_{LQ})$ space,
for different models of scalar and vector leptoquarks are presented
in Figure \ref{fig-lqres2}.
The 95\% CL exclusion limits on the leptoquark masses (i.e. largest mass 
values resulting in ${\cal P} \le 0.05$ for any value of $\lambda_{LQ}$) 
are included in Table \ref{tab-lqres1}.

\begin{figure}[tbp]
\centerline{\resizebox{\figwidth}{!}{%
  \includegraphics{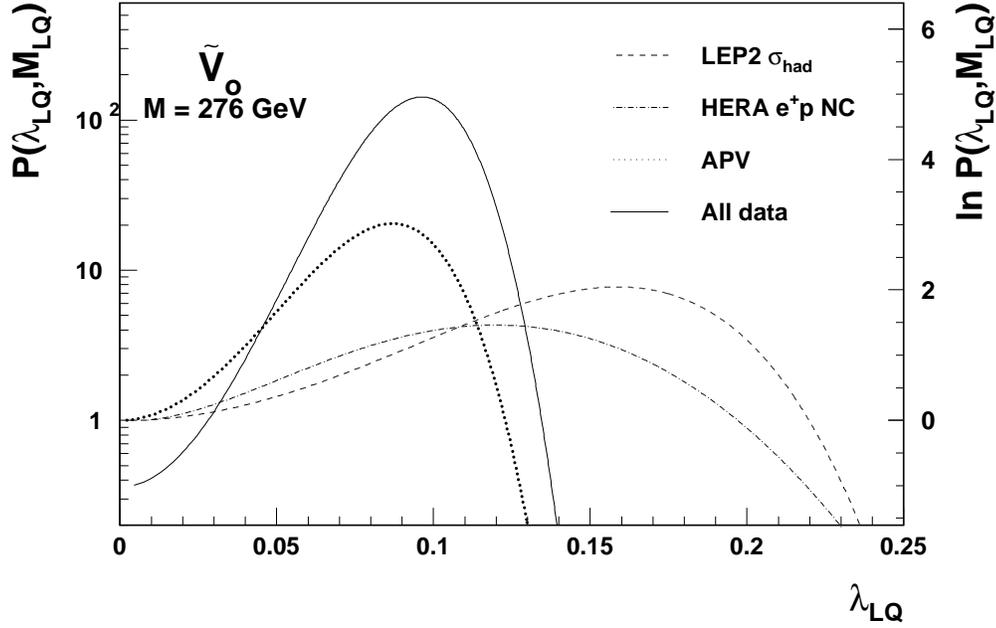}
}}
  \caption{ Contributions of different data sets 
            (as indicated on the plot) to the global
            probability functions ${\cal P}(\lambda_{LQ},M_{LQ})$ (left hand scale)
    and to the log-likelihood function  ${\ln \cal P}(\lambda_{LQ},M_{LQ})$
           (right hand scale), 
            for the $\tilde{V}_{\circ}$ model with $M_{LQ}=276$ GeV. }
  \label{fig-lqres6b}
\end{figure}

\begin{figure}[p]
\centerline{\resizebox{\figwidth}{!}{%
  \includegraphics{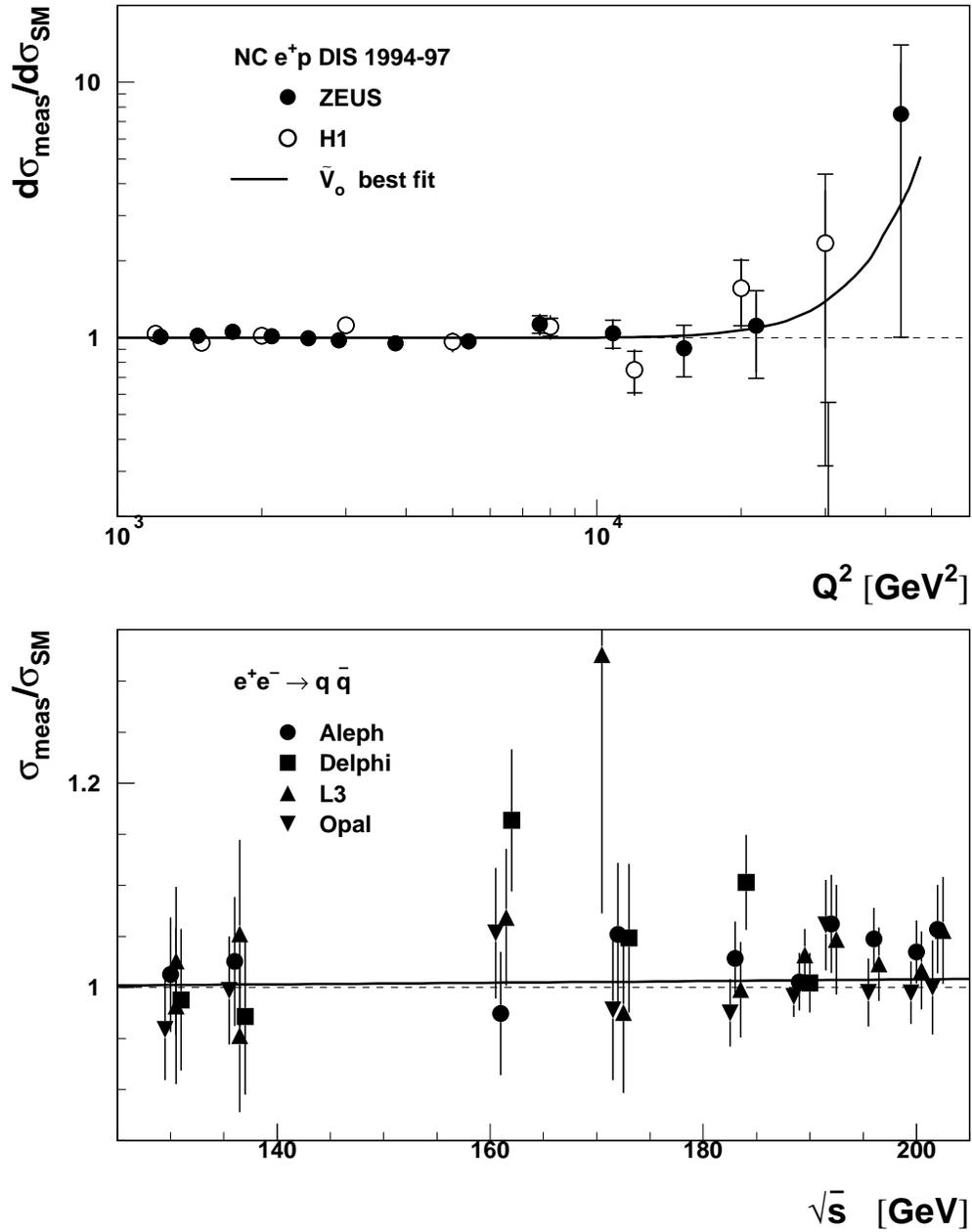}
}}
  \caption{
           Cross-section deviations from the Standard Model
           resulting from the fit of the  $\tilde{V_{\circ}}$ model
           (thick solid line) compared with 
           HERA NC $e^{+}p$ DIS cross-section results (upper plot)
           and LEP2 hadronic cross-section results (lower plot). }
  \label{fig-lqres4}
\end{figure}

\begin{figure}[p]
\centerline{\resizebox{\figwidth}{!}{%
  \includegraphics{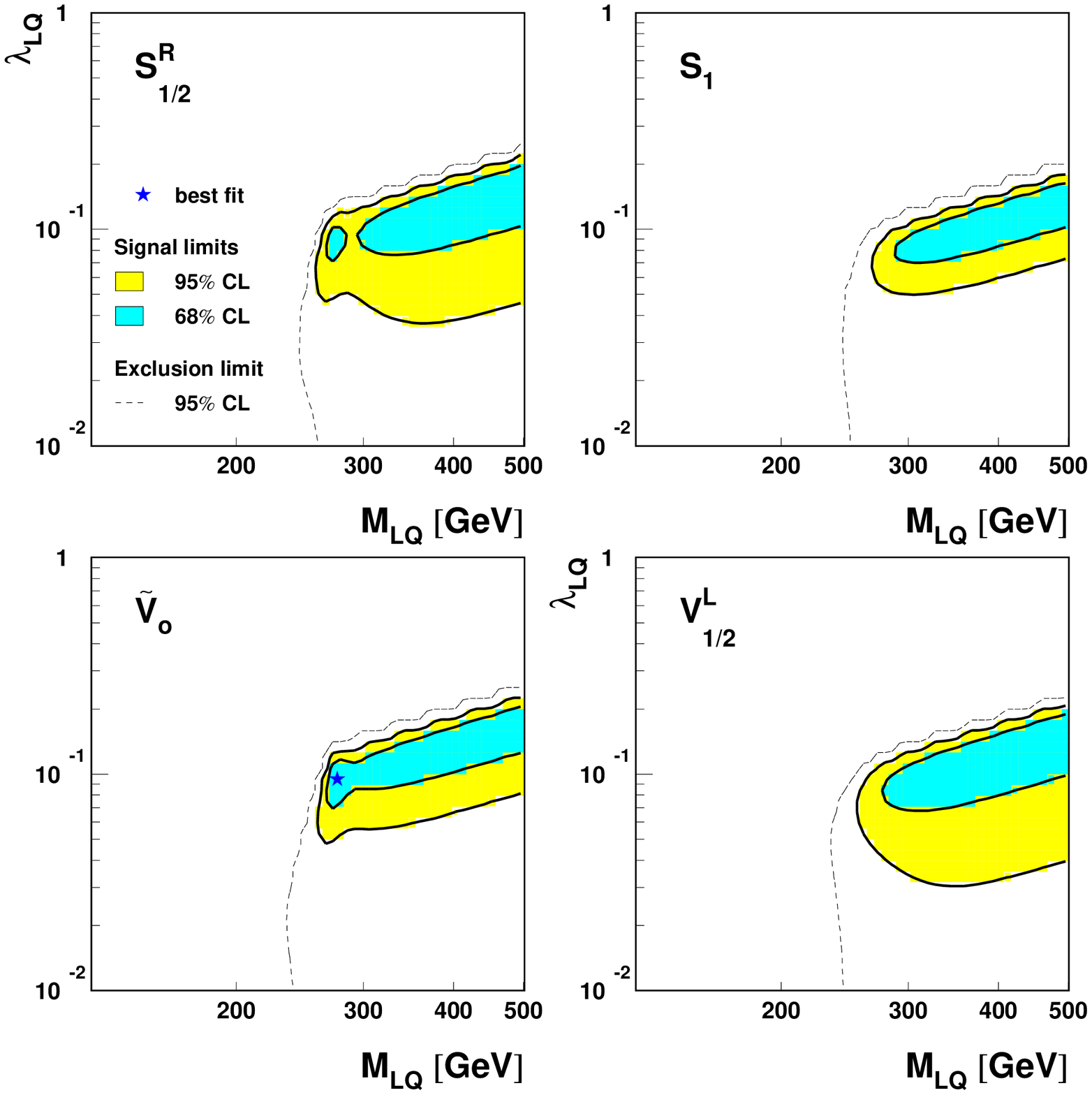}
}}
  \caption{ Signal limits on 68\% and 95\% CL for different leptoquark
            models, as indicated in the plot. Dashed lines indicate
            the 95\% CL exclusion limits. 
            For $\tilde{V}_{\circ}$ model
            a star indicates the best fit parameters.
            For other models the best fit is obtained in the contact
            interaction limit $M_{LQ} \rightarrow \infty$.}
  \label{fig-lqres3}
\end{figure}

The parameter values resulting in the best description
of the experimental data were also searched for for finite leptoquark
masses, varying $\lambda_{LQ}$ and $M_{LQ}$ independently.
Only for one leptoquark model an improvement has been obtained as compared with
the asymptotic solution.
For $M_{LQ}$=276 GeV and $\lambda_{LQ}$=0.095 the maximum probability
${\cal P}_{max}$=142 ($\ln {\cal P}$=5.0) is obtained for 
the $\tilde{V_{\circ}}$ model. This corresponds to about 3.1$\sigma$ 
deviation from the Standard Model. 
The effect is mainly due to the new APV measurement 
($\ln P$=3.0, 2.4$\sigma$ effect),
but is also supported by the excess of high-$Q^{2}$ NC $e^{+}p$ DIS events
at HERA ($\ln P$=1.4, 1.7$\sigma$ effect) and the LEP2 hadronic cross-section
measurements ($\ln P$=1.2, 1.5$\sigma$ effect).
For all HERA, LEP and low energy data the maximum probability 
turns out to be ${\cal P}_{max}$=367 (as compared to ${\cal P}_{max}$=122 
in the contact interaction limit). 
However, the value of  $\tilde{V_{\circ}}$ leptoquark mass 
of  $M_{LQ}$=276 GeV 
is already  strongly disfavoured by the negative
direct search results from the Tevatron ($P$=0.36, $\ln P$=-1.0).
Contributions of different data sets to the probability function
for the $\tilde{V}_{\circ}$ model with $M_{LQ}=276$ GeV
are presented in Figure \ref{fig-lqres6b}. 
Very good description of APV and HERA high-$Q^{2}$ data
is obtained for the fitted value of $\lambda_{LQ}$,
whereas LEP2 measurements again
suggest higher values of $\lambda_{LQ} \sim 0.16$.
The ratio of the predicted $e^{+}p$ cross-section at high $Q^{2}$
to the Standard Model cross-section is shown in Figure \ref{fig-lqres4}
together with the corresponding H1 \cite{h1pp} and ZEUS \cite{zeuspp} data. 
The hypothesis of the $\tilde{V_{\circ}}$ leptoquark production can describe 
the excess of events at highest $Q^{2}$ not affecting the perfect
agreement with the Standard Model at $Q^{2}\;<$ 10000 GeV$^{2}$. 
Also shown in Figure \ref{fig-lqres4} is the predicted deviation of the total
hadronic cross-section at LEP as a function of $\sqrt{s}$.
Best fit of the $\tilde{V}_{\circ}$ model results in the cross-section
increase at  highest $\sqrt{s}$ by about 1\%, 
which is consistent with available data.
From the fit of a two-dimensional Gaussian distribution in
the  close neighbourhood of the maximum of the probability function
${\cal P}(\lambda_{LQ},M_{LQ})$, the errors on the $\tilde{V_{\circ}}$ parameter
values were estimated:
\begin{eqnarray}
M_{LQ}  & = &  276 \pm 7 \; {\rm GeV} \nonumber \\
\lambda_{LQ} & = & 0.095 \pm 0.015 \;\; . \nonumber 
\end{eqnarray}
The local maximum of the probability function at $M_{LQ}$=274 GeV 
is also observed for the  $S^{R}_{1/2}$ model 
(${\cal P}$=20.3, as compared with ${\cal P}_{max}$=35.8 obtained in the high
leptoquark mass limit). 
This maximum is due to APV and HERA data, 
but is strongly suppressed by the Tevatron direct search 
results.\footnote{For the $S^{R}_{1/2}$ isospin doublet the combined
Tevatron 95\% CL limit is $M_{LQ} >$ 263 GeV, as compared with
the published limit of 242 GeV for single leptoquark production 
(see Section \ref{sec-tevdir}).}
Signal limits in the $(\lambda_{LQ},M_{LQ})$ space were studied 
for all leptoquark models which resulted in the description of 
the experimental data much better than the Standard Model
(${\cal P}_{max}>20$).
Best parameter values and 
estimated 95\% CL lower limits on the leptoquark masses are
summarised in Table \ref{tab-lqres3}.
In Figure \ref{fig-lqres3}, the signal limits at 68\% and 95\% CL 
are compared with exclusion limits  in the $(\lambda_{LQ},M_{LQ})$ space.
\begin{table}[tbp]
  \begin{center}
    \begin{tabular}{lcccccc}
      \hline\hline\hline\noalign{\smallskip}
Model & ${\cal P}_{max}$ & $\ln {\cal P}_{max}$ & 
$\left.\lambda_{LQ}\right|_{max}$  &   $\left.M_{LQ}\right|_{max}$  & 
$\left(\frac{\lambda_{LQ}}{M_{LQ}}\right)_{max}$  & 95\% CL limit \\
     &   &   &  & [GeV]  & ${\rm [TeV^{-1}]}$   &  on $M_{LQ}$ [GeV]  \\ 
\hline\hline\hline\noalign{\smallskip}
%
%
%
$S_{1/2}^R$  &   35.8  & 3.6 & 
  &   &  0.32 $\pm$ 0.06  &  
258  \\ 
$S_1$  &   367.  & 5.9 & 
  &   &  0.28 $\pm$ 0.04  &  
267  \\ 
$\tilde{V}_{\circ}$  &   142.  & 5.0 & 
0.095 $\pm$ 0.015 & 
276 $\pm$ 7 &     &  
259  \\ 
$V_{1/2}^L$  &   31.7  & 3.5 & 
  &   &  0.30 $\pm$ 0.06  &  
254  \\ 
\hline\hline\hline\noalign{\smallskip}
    \end{tabular}
  \end{center}
  \caption{Coupling $\left.\lambda_{LQ}\right|_{max}$ 
and mass $\left.M_{LQ}\right|_{max}$ values resulting 
in the best description of the experimental data 
 and the corresponding model probability ${\cal P}_{max}$  for
 different leptoquark models, as indicated in the table. 
 Also given are 95\% CL lower limits on the leptoquark mass
 (signal limits).
 Shown in the table are only those models which give much better 
 description of the experimental data than the Standard Model
 (${\cal P}_{max}>20$). 
 When the best description is obtained in the very high mass limit
 $\left(\frac{\lambda_{LQ}}{M_{LQ}}\right)_{max}$  is given.
}
  \label{tab-lqres3}
\end{table}

\clearpage
\section{Summary}
\label{sec-summary}

Data from HERA, LEP and the Tevatron, as well as from low energy  
experiments were used to constrain Yukawa couplings
for scalar and vector leptoquarks
in the Buchm\"uller-R\"uckl-Wyler effective model.
In the limit of very high leptoquark masses, constraints
on the coupling to mass ratio were studied using 
the contact-interaction approximation.
Some leptoquark models are found to describe the existing experimental data 
much better than the Standard Model.
The best description of the data is obtained for the $S_{1}$ model with
$M_{LQ}\gg$ 300 GeV and 
$\lambda_{LQ} / M_{LQ}\; =  \;0.28 \pm 0.04\; {\rm TeV^{-1}}$
and for the $\tilde{V}_{\circ}$ model with 
$M_{LQ} \; = \; 276 \pm 7$ GeV and
$\lambda_{LQ} \; = \; 0.095 \pm 0.015$. 
In both cases the increase of the global probability  
corresponds to more than 3$\sigma$ deviation from the Standard Model.
The effect is mainly due to the new data 
on atomic parity violation in cesium,
but is also supported by LEP2 hadronic cross-section results
and HERA NC $e^{+}p$ DIS (for the $\tilde{V}_{\circ}$ model) 
or low energy CC data (for the $S_{1}$ model).
Other data considered in this analysis are also in
good agreement with predictions these models.

If the observed $\tilde{V}_{\circ}$ signal is real it could
become visible in the new HERA $e^{+}p$ data, which are now
being collected at increased center-of-mass 
energy.\footnote{Since July 1999 HERA scatters 27.5 GeV positrons
on 920 GeV protons resulting in the center-of-mass energy
$\sqrt{s}$=318 GeV. The currently available HERA $e^{+}p$ data  
were collected 1994-97 with a proton beam energy of 820 GeV, corresponding
to $\sqrt{s}$=300 GeV.}

%
%
\section*{Acknowledgements}

I would like to thank all members of the Warsaw HEP group and 
of the ZEUS Collaboration for support, encouragement, many useful
comments and suggestions.
Special thanks are due to K.Doroba, J.Kalinowski and U.Katz 
for  many valuable comments to this paper.

This work has been partially supported by the Polish State Committee 
for Scientific Research (grant No. 2 P03B 035 17).

%
%


\end{document}